REVIEW

# Lymphocyte repertoire selection and intracellular self/not-self discrimination: historical overview

Donald R. Forsdyke

Correspondence: Dr. D. R. Forsdyke, Department of Biomedical and Molecular Sciences, Queen's University, Kingston, Ontario, Canada K7K3N6

Email: forsdyke@queensu.ca

**Immunological self/not-self discrimination is conventionally seen as an extracellular event, involving interactions been receptors on T cells pre-educated to discriminate, and peptides bound to major histocompatibility complex proteins (pMHCs). Mechanisms by which not-self peptides might first be sorted intracellularly to distinguish them from the vast excess of self-peptides have long been called for. Recent demonstrations of endogenous peptide-specific clustering of pMHCs on membrane rafts are indicative of intracellular enrichment before surface display. The clustering could follow the specific aggregation of a foreign protein that exceeded its solubility limit in the crowded intracellular environment. Predominantly entropy-driven, this homoaggregation would co-localize identical peptides, so facilitating their collective presentation. Concentrations of self-proteins are fine-tuned over evolutionary time to avoid this. Disparate observations, such as pyrexia, and female susceptibility to autoimmune disease, can be explained in terms of the need to cosegregate cognate pMHC complexes internally prior to extracellular display.**

**INTRODUCTION**

The predictions that the shaping of lymphocyte repertoires to meet future antigenic (e.g. viral) challenges requires both positive and negative selection[1] and that, for this purpose, 'promiscuous' tissue-restricted antigens (TRAs) are ectopically displayed in a central lymphoid organ (e.g. thymus),[2] have been confirmed in many laboratories.[3,4] Reactions between the receptors on T cells (TCRs) that have been 'educated' in this way, and peptides bound by major histocompatibility complex proteins (pMHCs) on antigen presenting cells (APCs), appear to provide sufficient self/not-self discrimination to prevent self-reactivity. However, recently there has been support for the prediction that this *extracellular* self/not-self discrimination is supplemented by prior *intracellular* discrimination.[5] Lu et al. reported in 2012[6] that, before pMHC display, "endogenous antigen processing generates intracellular clusters of class I molecules segregated on the basis of their peptide cargo." While confirming the high specificity of this colocation of cognate pMHC complexes, Ferez et al. in 2014[7] regretted that the mechanism of the "preferential loading of newly synthesized MHC class 1 loading with viral peptides," remained problematic.

Indeed these, and many other spectacular advances, bring to light as many problems as they solve. Both centrally and peripherally the TCR specificity of T cells is set to anticipate future pathogens, but how did this evolve? Why is there less redundancy of TCR binding (less cross-reactivity) than first thought? Why are there 'coexpression groups' of TRAs, and why is there mosaic, not ubiquitous, TRA expression in medullary thymic epithelial cells (mTECs)? Why do the self-peptides isolated from MHC proteins usually correspond to large and abundant proteins? Is it better to think in terms of TCR affinity or avidity?

Then there are the much broader questions that are not usually dealt with in this context. When the crunch comes, is it more important to respond to foreign or prevent reaction with self? Why are females more prone to autoimmune disease? What is the adaptive value, if any, of pyrexia? Can T lymphocyte reactivity with polyclonal mitogens (e.g. lectins) tell us anything about reactivity with specific antigens? Why in infectious and autoimmune diseases are there changes in plasma such that erythrocytes aggregate into rouleaux that sediment rapidly in isolated blood samples?

The latter, known clinically as the increased ESR (erythrocyte sedimentation rate),[8] focused attention on 'macromolecular crowding' and the many clinical disorders associated with specific extracellular or intracellular protein aggregates. The present paper relates a two decade old hypothesis on the role of protein aggregation in intracellular self/not-self discrimination[5] to new developments in our understanding of pMHC presentation and thymic function. While still far from definitive, that the proposed mechanism appears to unify the many disparate problems listed above, suggests review is timely.

**THE NEAR-SELF REPERTOIRE**

A two signal hypothesis[9,10] postulated a distinction, at the level of individual immunologically-competent cells, between those specific antigenic signals that activate, and those that inhibit. Although the division is not absolute, a comparable cell fate duality maps for T lymphocytes to two thymic locations. Activation signals associated with positive repertoire selection locate mainly to the cortex, and inhibitory signals associated with negative repertoire selection locate mainly to the medulla.[4] After endowment, in a quasi-random manner, with a wide range of specificities, developing thymocytes proceed from cortex to medulla, running sequentially the gauntlets of positive and negative selection. The few that 'audition' successfully enter the peripheral lymphoid system.

Ideally this somatic 'education' process would remove self-reactivity and anticipate reactivity to foreign 'not-self' antigens. However, given that specific foreign challenges are often from microorganisms with higher mutation rates than those of their hosts, such anticipation would seem in vain. Similarly, the notion that, over evolutionary time, there is germ-line selection of prospective hosts with MHCs preadapted for such challenges, is increasingly seen as unlikely.[11,12] Thus, the view that there cannot be germ-line information for anticipating the specific immunological challenges of future generations – i.e. they must start with neutral 'blank slates' – has grown more secure. Nevertheless, the possibility that immunological repertoires might become skewed during *somatic* time to the advantage of hosts emerged when the process was examined from the perspective of pathogens.[1]

A microbe that could, in one step, mutate one of its antigens from a form that was not-self with respect to its host, to a form that was self with respect to its host, would have largely overcome the host's immune defences with respect to that antigen. It could then exploit the 'holes' in the repertoire that had been created by the host's prior elimination of self-reacting lymphocytes. However, mutation is generally a stepwise process. If a microbe (not-self), by mutating a step towards self along the path from not-self to self, could secure a selective advantage, then the mutant form would come to dominate the microbe population. If a microbe from this mutant population, by mutating a further step along the path, secured a further advantage, then this new mutant form would, in turn, come to dominate the population. Thus, an average member of the microbe population would progressively become better adapted, to the detriment of the host. This supposes that progressive mutation along the not-self-to-self path would be increasingly advantageous to the microbe. However, the advantage would be lost if, as it mutated closer to host-self, the microbe encountered progressively stiffer host defences. Thus, positive selection of lymphocytes for specificities that were very close to, but not quite, anti-self – that is anti-'near-self' specificities – could be an important host adaptation.[1]

Indeed, it is now recognized, both that positive selection of T cells of intermediate affinity for 'near-self' shapes immunological repertoires,[3,13–16] and that the repertoires so-skewed can achieve selective high affinity targeting of pathogens.[17] In 2013 Mandl et al.[18] observed: "TCRs able to bind self pMHC well (but below the negative selection threshold) also bind especially well to foreign pMHC and hence … positive selection ensures that T cells most useful for host defense against pathogens are selected from a diverse initial repertoire to populate the peripheral T cell pool." Thus, "the *raison d'être* of positive selection … is to bias T cell selection towards strongly self-reactive clones that are endowed with a homeostatic advantage and a head start in anti-pathogen responses".[4] Cancro and Kearney[19] have written similarly on the 'education' of B lymphocytes whose receptors share "parallel strategies of antigen recognition" with T cells.[11]

In this way, by anticipating their mutational strategies, a host *can*, albeit indirectly, have prior knowledge of pathogens. Such reasoning questions the notion that heterozygote advantage drives the evolution of MHC polymorphism.[20,21] In 2004 Borghans et al.[22] calculated that: "Host-pathogen coevolution … can easily account for realistic polymorphisms of even more than 50 alleles per MHC locus." Furthermore, focusing on 'near-self' (i.e. limited cross-reactivity) greatly reduces the need for TCR binding degeneracy to cope with the universe of potential peptide challengers.[23,24]

The somatic 'education' of lymphocytes is not confined to central lymphoid organs, but is ongoing.[25] Indeed, successful 'auditioners' that escape from central to peripheral lymphoid tissues are kept tuned to near-self and maintained in a state of constant readiness through 'tonic' low affinity interactions with near-self antigens that deliver survival signals, but do not initiate proliferation in the absence of general homeostatic signals.[26] Thus, there is "a link between thymic pMHC experience and mature T cell homeostasis".[4] A lymphocyte 'preactivated' in this way is poised to respond. The onus is then on APCs to present appropriate high affinity pMHCs when circumstances so warrant.

In addition to the thymic cortex/medulla locational duality, the modes of antigen presentation by different types of antigen-presenting cells (APC) in these two locations differ profoundly. We here review how an understanding of the latter duality can relate both to two signal ideas,[1,9] and to the postulate of a *two-step* self/not-self discrimination processes – first an intracellular discrimination, and then an extracellular discrimination mediated by 'educated' T lymphocytes.[5]

## SELF PEPTIDES AS DISTRACTORS

Amidst a sea of near-self pMHCs for which they have intermediate affinity, peripheral αβ TCR lymphocytes are kept in constant readiness for the rarer high affinity complexes that should focus them to the displaying cell, be it a 'professional' APC with pMHC directed to CD4 and CD8 coreceptor T cells, or another cell type with pMHC directed to CD8 T cells. The need for an urgent response suggested by the state of constant readiness, indicates advantages in APC mechanisms that could focus attention on newly arising high affinity pMHCs derived from not-self peptides, so avoiding distraction by the large excess of intermediate affinity pMHCs derived from self-peptides. This made attractive, hypotheses that peptides from freshly synthesized partial or complete proteins might somehow be selectively channeled to pMHCs. Thus, views that peptides were formed from incompletely synthesized proteins – 'defective ribosome products' (DRiPs) or 'pioneer translation products' (PTPs) – gained much attention.[27,28]

However, two recent advances make timely the recalling of an alternative mechanism.[5] First, views that peptides are formed from DRiPs or PTPs are now contested by evidence that peptides in pMHC complexes derive from native, fully synthesized, properly folded, proteins. These mature proteins can be degraded to peptides at any time after synthesis and are not necessarily 'retirees' that are part of normal protein turnover.[29–31] Indeed, from studies of an intracellular parasite that secretes only mature proteins into host cytosol, Wolf and Princiotta[32] conclude that "presentation efficiency may be higher for proteins that enter the cellular pool when compared with those processed in a near cotranslational manner, such as endogenously synthesized … DRiPs." Second, there is a better understanding of how APC in the thymic cortex (cortical thymic epithelial cells; cTECs), differ in peptide-generation mechanisms from various APC in the thymic medulla (including medullary thymic epithelial cells; mTECs). Medullary peptide-generation mechanisms more closely resemble those found in peripheral APCs.[4] Can the division of labour – cTECs relating to positive selection and mTECs relating to negative selection – tell us something about peptide sorting mechanisms and the distinction between signals that activate, and those that inhibit? We first consider how a signal that activates at one time may inhibit at another, so the distinction must be seen in context.

**ANTIGEN DOSE OPTIMUM SHIFTS**

Autoimmune diseases, reflecting a failure of negative selection, are usually of slowly increasing severity, whereas attacks by microbial pathogens, requiring positive selection, often need urgent responses. Assuming a duality related to antigen dosage, and that dosage would be low at early time points, it would seem more likely *a priori* that a low dose of an antigen would suffice for positive selection of responding lymphocytes, whereas a high dosage would be required for their negative selection. Building on the observations that (i) an effective form of immunological tolerance is, indeed, antigen dose-dependent – high doses being tolerogenic,[33] and (ii) activation of cultured lymphocytes by a polyclonal mitogen is an all-or-none ('digital') phenomenon,[34] a two signal hypothesis[9] proposed that low doses would provide the signal to a lymphocyte for an immune response (positive selection) and high doses would tolerize (negative selection; clonal deletion). Indeed, it was found, that high concentrations of a polyclonal mitogen deleted cultured lymphocytes, that deletion was complement-dependent, and that it involved cross-linking of cell-borne receptors.[35,36] There is now evidence, for both CD4 and CD8 T cells, that one pMHC can activate and that two or more pMHCs can kill.[37,38] However apoptosis, rather than complement, appears as the agency.

   Urgency of activation relative to inhibition is supported by in vitro murine antigen dose-response studies by Alexander-Miller et al..[39] At early time-points the function (cytotoxic activity) of T cells progressively increases with antigen dose. However, at later time-points the higher antigen dosages are inhibitory. Thus a concentration that appears optimum at late time points is less than the optimum as assessed at earlier time points. Similar optimum shifts are seen with lymphocytes cultured with varying doses of antigens or polyclonal mitogens.[40,41] Furthermore, the late-onset high dose inhibition begins at lower dosages with high avidity antigens.[39,42] Optimum shifts can also appear *in vivo*. In 1964 Mitchison found that high dose tolerance was of late onset following an initial immunogenic phase.[33] In 1996 Liblau et al. found that, whereas thymocyte tolerance induced by high antigen dosage was rapid, peripheral tolerance followed a period of lymphocyte activation.[43]

**THYMIC POSITIVE AND NEGATIVE SELECTION**

Following Occam's razor, in 1999 Barton and Rudensky[44] thought "it would be reasonable for the immune system to evaluate T cells during development on the basis of the rules of recognition that are required in the periphery." Thus, by analogy with the signaling duality in peripheral lymphocytes (see above), thymic positive selection might require low antigen dosage, and thymic negative selection might require high antigen dosage. However, the thymic cortex is distinct from the periphery in that it is a site for generation of wide TCR diversity, with apoptotic loss 'by neglect' of those cells whose TCRs do not achieve a minimum level of affinity for the pMHCs displayed by cTECs.  Only when members of such a 'preselection repertoire' have been generated can they be subjected to positive selection for cells with TCRs of moderate and high affinity for self pMHCs. In this process they may be activated to some degree,[45] but proliferation ceases.[46] Indeed, since many cells are to be negatively selected, proliferation would seem unnecessary at this stage.

Nevertheless, the activation might follow the 'rules of recognition' operative in the periphery. But whereas mature peripheral lymphocytes can be activated by polyclonal activators (lectins), thymocytes cannot.[47] Furthermore, activation of the 'preselection repertoire' is indifferent to the CD4 or CD8 nature of coreceptors, and only needs to be permissive or non-permissive – above a critical threshold (below which 'neglect' would be in play) activation needs to depend neither on affinity for, nor the actual dosage of, the activating pMHC. In other words, there is wide affinity window.[48] Thus, there would be no need for cTECs to generate large quantities of individual proteins in order to generate peptides.[44] Indeed, 'preselection repertoire' diversity might be maximized if the resources for protein production were devoted to producing many proteins, rather than large quantities of individual proteins. In this circumstance, as suggested by Linsk et al. in 1989,[2] a low level of whole genome 'transcriptional noise' – otherwise 'illegitimate,' 'ectopic,' or 'leaky' transcription – might suffice to produce sufficient quantities of each protein for positive selection.[49]

On the other hand, thymic negative selection requires high concentrations of individual proteins. A well-studied example is proinsulin, where a polymorphic variant associated with high thymic expression spares humans from type 1 diabetes.[50] Likewise for rodents Kyewski and Klein[51] note "the … exquisite sensitivity of the central tolerance process to moderate quantitative variations in TRA expression. Thus, subtle differences in the range of two- to fourfold in intrathymic expression of auto-antigens as obtained in mice with defined copy numbers of pancreatic or nervous system-specific antigens can modulate the susceptibility to autoimmunity."

A need for high concentrations implies that each mTEC would limit the diversification of its proteins, so that its resources for protein production could better produce large quantities of a few proteins from which corresponding pMHCs would be produced.[2] This would be consistent with a dramatic difference between cTECs and mTECS – namely the mosaic location within distinctive medullary sectors of the specific pMHC-bearing mTECs that 'promiscuously' synthesize different TRAs. It is estimated that, at given time, only 1–3% of mTECs produce a particular antigen.[52] While currently there is much concern about differences between cTECS and mTECs in the *qualitative* characteristics of the peptides produced – cTECs generate a unique

"MHC ligandome"[4] – this could be merely ancillary to the need to optimize these two different roles.

Nevertheless, one qualitative characteristic of a mature protein from which peptides are derived is important – its function. Once synthesized within the thymus, a promiscuous gene product is not needed for its function, but for its components. A need for high quantities raises the prospect of a degree of ectopic function that could endanger the cell. For this reason, it would be predicted, first, that there would be an intracellular mechanism to limit function (e.g. rapid inactivation and dismemberment following synthesis; see below), and second, that mTECs would have a shorter lifespan than cTECs,[53] so there would be no medullary equivalent of cortical 'nurse cells'.[4]

A similarity between medullary and peripheral mechanisms of negative selection would seem necessary to avoid central negative selection generating unduly large 'holes' in the final repertoire.[23] If the probability of a particular self-peptide ever being presented peripherally were remote, then a medullary state under which that pMHC might invoke negative selection would be superfluous.[54] Indeed, Kyewski and Haskins[55] note that: "cTECs represent a unique APC type in the body … . In contrast, the two major medullary APC types, DCs and mTECs, very much resemble peripheral APCs." Thus it would seem appropriate to look to peripheral APCs for guidance on mechanisms that might exist in mTECs.

However, when contrasting medullary negative selection of developing thymocytes with the negative selection of mature peripheral lymphocytes, another difference emerges. Despite possible shifts in optimum dosage (see above), a peripheral lymphocyte interacting with pMHC can be deemed as held within a narrow time-window within which there must be a decision between positive activation and negative selection. On the other hand, for cells of the post-positive selection thymocyte repertoire that enter the medulla there is only one fail/pass decision to be made – either be negatively selected or not. This would seem to require *some* pMHC dose on mTECs, but not necessarily a high pMHC dose. We are then returned to the "MHC ligandome" dilemma referred to above, which would seem better resolved by ascribing differences in cortical and medullary selections to *qualitative* differences in peptides.[4] However, this problem can be addressed if we ascribe the need for high antigen dosage to *a critical intracellular event that surface pMHC dosage comes to reflect*.

**INTRACELLULAR SELF/NOT-SELF DISCRIMINATION**

Given that peripheral APCs can present both self pMHC and non-self pMHCs, it is easy to suppose that the key to self/not-self discrimination is the central 'educational' process that creates a T lymphocyte repertoire that does not respond immunologically to self, but does respond to not-self. However, a mechanism for over-riding distracting self pMHCs in favour of a rare non-self pMHC, would seem advantageous.[20,56] Thus, the processing of self pMHCs responsible for the 'tonic' stimulation of T-cells in the periphery should somehow differ, either absolutely or in degree, from the processing of pMHCs derived from proteins deemed not-self.

There is now evidence that, prior to appearance at the APC surface, there is *intracellular* clustering of pMHC complexes with the *same* peptide cargo (i.e. cognate pMHC complexes cosegregate internally). Lu et al.[6] regard this as "our most important finding," and Ferez et al.[7] find as the "more important implication" that the peptide-specific clustering "is indicative of intracellular enrichment." Noting that the "exact mechanism causing the cognate pMHC enrichment remains unclear," Ferez et al. suggest that "it may result from the burst of viral protein expression that peaks a few hours post-infection. This burst could lead to a preferential loading of newly synthesized MHC class I complexes with viral peptides." In other words, a virus that expresses its proteins in a burst automatically identifies itself to the host as 'not-self'!

However, the proposed "enrichment" of "identical peptides" raises the possibility of a special mechanism for their intracellular *colocation* to separate them from other more diffusely distributed peptides (i.e. the possibility of a specific *intracellular discriminatory process* that is not so easily subverted). This colocation mechanism could act at the level of diffusely distributed individual pMHC complexes, or could occur earlier. There is only one copy of a particular peptide within a protein, which often displays only that one peptide. So the colocation could first occur at the level of the individual peptides once they had each been released from diffusely distributed donor protein molecules. Alternatively, the colocation could first occur at the level of the donor proteins, so that the released peptides would then be available in close proximity for formation of cognate pMHC clusters. A mechanism consistent with the latter alternative[5] is discussed below.

If there were an intracellular self/not-self discrimination process, it could have first evolved at a primitive unicellular level where it would have served to rapidly limit the function of components of an intracellular pathogen and/or to trigger host apoptosis. The latter would altruistically militate against spread of the pathogen to neighbouring members of the species that shared host genes.[57] Even supposing that the 'hand of evolution' had conjured up such a mechanism, it alone would not have sufficed in multicellular organisms where something equivalent to a pMHC display would have served to inform an entire organism that one of it cells contained something designated as 'not-self.' The primitive T cells that recognized that cell, would not only kill the cell, but would be stimulated to divide, so producing more T cells of the same specificity, which could then, both seek other cells that displayed the same foreign pMHC, and establish immunological memory. Thus, immediate apoptosis triggered by an intracellular self/not-self discrimination event would seem counter-productive.

## INTRACELLULAR AGGREGATION HYPOTHESIS

Building on studies of the specificity of the aggregation of erythrocytes into rouleaux by proteins and other agents that *do not directly interact* with the erythrocytes,[8,58] the problem of intracellular self/not-self discrimination in the 'crowded' intracellular environment was addressed in the decade after the discovery of the association of peptides with MHC proteins.[20,59,60] Cytosolic proteins are deemed to exert a *collective* pressure tending to make individual protein species coaggregate (self-assemble) when their concentrations exceed their individual solubility limits. These concentrations have been fine-tuned to the concentrations of their evolutionary 'fellow travellers,' so as not to exceed these limits. Not-self proteins more readily 'trip' this intracellular surveillance system because their concentrations have not been so fine-tuned.[5] When a virus infects a cell, *at best*, this process results in pMHC display and the destruction of the cell; *at least*, the virus is forced to mutate in order to avoid identification as not-self, and the mutated form may proliferate less well.[61,62]

When macromolecules in solution reach a critical concentration it becomes energetically more favourable for them to aggregate, like-with-like, than to remain in simple solution. The aggregation involves liberation of bound water and an increase in entropy. Being primarily entropy driven, the aggregation is promoted by an increase in temperature.[63] That the crowded cytosol constitutes an environment which readily drives proteins out of solution when they exceed individual concentration thresholds, is well recognized from the difficulties encountered when trying to over-express proteins within foreign cytosols using expression vectors. Lowering culture temperature is a common strategy to overcome this. Likewise, increasing an organism's temperature (pyrexia) can be seen as a short-term strategy for increasing the probability of aggregation when the adaptive value of recognizing foreign may be greater than that of avoiding self-reactivity. Furthermore, an organism retains the option of 'declaring' one of its own self proteins 'foreign' should its sequence or expression change (due to mutation).

The exquisite like-with-like specificity that intracellular aggregation can achieve is now well documented.[64] Such homoaggregation often results in loss of function. But the scale of the aggregation in APC envisaged here ('microaggregation'), suggests experimental detection would not be easy. However, macroscopic aggregates ('inclusion bodies') are a feature of various clinical conditions (e.g. Parkinson's disease). Although prior homoaggregation occurs independently of the formation of inclusion bodies, which can contain various proteins, their presence provides a measure of the predisposing aggregation process. If factors promoting or impeding the formation of such macro-aggregates similarly affect pMHC displays in infected cells, this can be construed as indirectly supporting the view that micro-aggregation is a necessary stepping stone leading to those displays. Rather than repeat previous arguments for the aggregation hypothesis,[5,21] the present paper draws attention to recent work supporting the stepping stone viewpoint that aggregation first marks a protein as 'not-self,' and this is an essential prerequisite for the display of one or more of its peptides as pMHCs.

**NEW EVIDENCE ON AGGREGATION**

Predicting tight control of intracellular protein concentrations, the aggregation hypothesis focused on the evolution of sex chromosome dosage compensation.[60] Evidence for a chromosomal, rather than hormonal, basis for sex differences in the incidence of autoimmune diseases,[65] now supports the view that failure of human females to adequately turn-off one X chromosome, so increasing aggregation pressure, could explain their marked predisposition to autoimmune diseases.[66] Consistent with this, males with chromosomal anomalies such that there are two X chromosomes, are also predisposed.[67]

Since one of the roles of heat shock proteins (HSPs) is to chaperone intracellular proteins and *reverse* their aggregation, the existence of a class of inducible HSPs that would *promote* aggregation was postulated.[68] Their induction would associate with the pyrexia accompanying an antigenic challenge, and their experimental or therapeutic elimination would decrease aggregation and hence could be of value when studying or treating autoimmune diseases. Vitiligo (patches of white skin) is a positive prognostic factor in patients with melanoma tumours, indicating immune attack directed both against tumor antigens and those of pigmented skin cells. This T cell-mediated 'collateral damage' of skin cells is decreased when inducible HSP70 is inactivated, and increased when it is overexpressed.[69] Consistent with this, overturning long-held contrary views, formation of the intracellular aggregates in Huntington's disease is now reported to be *favoured* by the heat-shock response that should associate with pyrexia.[70]

Another prediction of the aggregation hypothesis was that proteins should be under evolutionary constraint not only to retain specific function, but also to retain solubility and lifespan (required for their collective function – the exerting of aggregation pressure). Thus, organisms synthesizing proteins with mutations adversely affecting the latter properties, would be selected against.[71] Molecular properties such as isoelectric point and size (which affect ability to aggregate) would be expected to vary less than predicted on the basis of known amino acid substitution rates. This was observed as low inter-species variation in two dimensional gel electrophoretic analyses of proteins,[72] and is consistent with work showing that variance in mRNA concentrations is much greater than the variance in concentrations of the corresponding proteins.[73] And from bioinformatic analyses, in 2012 Hoof et al.[74] concluded that: "protein abundance carries more information for the prediction of protein sampling [for MHC presentation as peptides] than transcript levels do."

The hypothesis that protein concentrations have evolved to contribute, and to respond, to the aggregation pressure exerted collectively by intracellular proteins, predicts that excess of one protein will cause differential aggregation of others, thus exerting pleiotropic effects (e.g changes in functions not necessarily related to that of the original protein[75]). Furthermore, aggregation of intact proteins is favoured when they are *large* and/or *abundant*, the latter being likely to correlate with gene expression level in terms of mRNA concentrations. In a survey of pMHCs, Fortier et al.[76] found that most peptides are derived from abundant mRNAs. And peptides are disproportionately presented from large proteins.[74,77]

Another prediction is that proteins contributing most to macromolecular crowding will be conserved by virtue of this property. This means that highly expressed proteins are likely to be more conserved (evolve slower) than lowly expressed proteins. The conservation affects both protein surface residues so that reactivity with the surfaces of other proteins is decreased ('misinteraction avoidance'), and protein cores ('misfolding avoidance'). This highly significant negative correlation between the expression level of a protein and its rate of evolution (the 'E-R anticorrelation') has been most studied in bacteria and yeast, but there is suggestive evidence for its generality.[78,79] There is also a highly significant negative correlation between expression level and the finely tuned propensity to aggregate.[80]

If differential aggregation is the mechanism by which intracellular self/not-self discrimination *initiates*, it would be predicted that the generation of aggregrates would *precede* any covalent modifications that might lead to peptide generation by way of the proteasome-ubiquitin system or other channels. Although there is no direct evidence on this, recent studies in pathological aggregation systems indicate that ubiquitination or phosphorylation occur *after* aggregation.[81,82] This suggests a necessity for prior aggregation in order that these downstream events can occur.

Regarding non-allelic genes, by limiting the range of proteins ectopically expressed at any one time, an mTEC decreases the chance of the co-aggregation of two proteins that would normally not be in the same tissue environment together. This would be part of the strategy of minimizing the number and size of potential 'holes' in the T cell repertoire.[83] Indeed, there is evidence that discrete sub-sets of genes ('coexpression groups') with distinctive chromosomal locations are expressed at different time-points.[4] These may have been selected over evolutionary time because they do not coaggregate. Whereas most autosomal genes are biallelically expressed, some autosomal genes, like female X chromosomal genes, are monoallelically expressed in a random manner. Biallelic or monoallelic expression as random alternatives in mTECs[84] might facilitate thymocyte selection (or deletion) in heterozygotes if aggregation were impeded (or advanced) by the heterozygosity.

## THE AVIDITY INTERPRETATION

Evidence for this proposed designation of a protein as not-self by specific intracellular aggregation is currently indirect. But assuming it to occur, there should then be a corresponding pMHC display. The number of pMHCs displayed should be reflective of the intracellular concentration of the peptide-donor protein within a peripheral APC. Given that there is a sufficient concentration in the first place to permit the aggregation, what is it about the subsequent display that reflects an *intermediate* concentration of a protein corresponding to a high affinity pMHC (hence requiring an immune response)? And what is it about the display that reflects a *high* concentration of a protein corresponding to a high affinity pMHC (hence requiring a tolerogenic response)?

Following the minimal two signal postulate,[9] one pMHC should be stimulatory and multiple pMHCs would be inhibitory.[85,86] This model is among those classified in 2013 by Bains et al.[87] as "p – sum," where "T cells continually re-assess fate decisions on the basis of multiple summed proximal signals from TCR-pMHC interactions." An additional postulate was that the inhibition would require that the signaling receptors be in close proximity.[9] Thus, a model for the shaping of lymphocyte repertoires by both positive and negative selection employed the term 'avidity,' rather than 'affinity'.[1] It was held that there must be "close receptor-determinant interaction for a discrete period of time" – a function that relates "to the chemical affinity of the receptor for the determinant." Furthermore, the "stimulation of the cell to initiate an immune response follows reaction of antigenic determinants with a limited number of cell-borne receptors," but "at high determinant concentrations … there is increased probability of the simultaneous occurrence of two reactions in close proximity at the lymphocyte surface between cell-borne receptors and antigenic determinants." Indeed, for thymocytes, Suzuki et al.[88] proposed that low degrees of cross-linking were required for positive selection and more extensive cross-linking was inhibitory. In 1997 Girao et al. noted:[89]

> "The critical parameter determining the developmental fate of thymocytes is the avidity of interaction between thymocyte TCR and peptide/MHC complexes of thymic stromal cells; low avidity interaction results in no selection and the death of thymocyte by programmed death, intermediate avidity results in the rescue of thymocytes from programmed death and positive selection, and high avidity results in negative selection. One of the original caveats of this model is that there may be upper and lower limits to the intrinsic affinity of a peptide that may prevent it from triggering the positive or negative selection of a thymocyte, irrespective of the density on thymic stromal cells. For example, a peptide may have such high affinity for thymocyte TCR that even at very low density on thymic stromal cells the avidity of interaction is too high for positive selection and only induces negative selection; conversely, a peptide may have such a low affinity for thymocyte TCR that even at maximum density on thymic stromal cells it is incapable of generating sufficient avidity for positive or negative selection."

Their experiments lead them to conclude that "the overall avidity, and not solely the affinity of TCR-peptide/MHC interaction, determines the developmental fate of thymocytes."

But in a later consideration of "hypotheses to resolve the selection paradox," Klein et al.[90] distinguished models "based on the avidity or the affinity of the TCR–peptide–MHC interaction," with the severe caveat: "Although the two models are frequently used synonymously, they are based on distinct assumptions. The avidity model predicts that the quantity of a given peptide–MHC complex expressed by cTECs dictates whether a thymocyte expressing an interacting TCR will be positively selected or deleted, whereas the affinity model instead postulates a crucial role of the quality of the individual TCR–peptide–MHC interaction." However, recent progress in the field seems to have led the authors to refer to "affinity and/or avidity," and to write simply of "strong interactions" and "weak interactions".[4]

Indeed, there is now considerable evidence for cognate pMHC coexpression in membrane nanoclusters that crosslink TCRs during interaction between APC and a T cell.[91,92] This tends to be interpreted as a means of achieving more efficient early T cell *activation*. Ferez et al. note:[7]

> "The detection of clusters of MHC molecules presenting identical viral peptides upon virus infection provides a solution to the paradox between the long-known need for multivalency of experimental, activating TCR ligands … and the notion that there is only a very small probability that a few identical pMHC complexes within a sea of irrelevant pMHC complexes will ever get close enough to engage the TCR in a multivalent fashion. Second, they support a model where the interaction between the TCR and its pMHC ligands is multivalent, providing a mechanism whereby cooperation between low-affinity interactions leads to an increased apparent affinity and thus high sensitivity."

Nevertheless, possible late-onset inhibitory effects, reflecting shifting dose-response relationships (see above), were not excluded. Since the degree of cross-linking would be less important in the cortex, cTECs would require much lower pMHC levels than mTECs, as is observed.[13] It would also be predicted that formation of membrane nanoclusters would be less necessary in cTECs. Consistent with this, the expression of a marker for membrane lipid rafts was not maximal until a late stage of positive thymic selection.[93]

**CONCLUDING REMARKS**

Of four consecutive hypotheses – two signal,[9] positive selection,[1] protein-based intracellular self/not-self discrimination,[5] and 'antibody RNA'-based intracellular self/not-self discrimination[94] – the first two, albeit in various guises, are now widely accepted. Indeed, noting that "somewhat paradoxically, recognition of self can elicit diametrically opposed outcomes," Klein et al.[4] declare: "The classical affinity model of thymocyte selection offers an attractive conceptual framework to resolve this apparent contradiction." However, they regret that "it does not take into account the fact that positive selection and negative selection mainly occur in discrete thymic microenvironments … the cortex and the medulla." The present paper has addressed this duality in terms of a differential protein aggregation mechanism for self/not-self discrimination that draws parallels with the aggregation phenomena found in various diseases and is informed by studies of lectin-lymphocyte interaction. The paper has considered neither the nature of 'third signals' that entice some CD4+ T cells to adopt other roles, nor the roles of non-αβ T cells, nor the putative 'antibody' role of RNA.[57,61,83,95] Nevertheless, it is hoped that the proposed explanations for numerous disparate phenomena will fruitfully guide future experimentation.

## ACKNOWLDEGEMENTS

---

1 Forsdyke DR. Further implications of a theory of immunity. *J Theor Biol* 1975; **52**: 187-198.

2 Linsk R, Gottesman M, Pernis B. Are tissues a patch quilt of ectopic gene expression? *Science* 1989; **246**: 261.

3 Forsdyke DR. Immunology (1955-1975): The natural selection theory, the two signal hypothesis and positive repertoire selection. *J Hist Biol* 2012; **45**: 139-161.

4 Klein L, Kyewski B, Allen PM, Hogquist KA. Positive and negative selection of the T cell repertoire: what thymocytes see (and don't see). *Nature Rev Immunol* 2014; **14**: 377-391.

5 Forsdyke DR. Entropy-driven protein self-aggregation as the basis for self/not-self discrimination in the crowded cytosol. *J Biol Sys* 1995; **3**: 273-287.

6 Lu X, Gibbs JS, Hickman HD, David A, Dolan BP, Jin Y, et al. Endogenous viral antigen processing generates peptide-specific MHC class I cell-surface clusters. *Proc Natl Acad Sci USA* 2012; **109**: 15407-15412.

7 Ferez M, Castro M, Alarcon B, van Santen HM. Cognate peptide–MHC complexes are expressed as tightly apposed nanoclusters in virus-infected cells to allow TCR crosslinking. *J Immunol* 2014; **192**: 52-58.

8 Forsdyke DR, Ford PM. Rouleau formation as a measure of the phase-separating ability of plasma. *J Theor Biol* 1983; **103**: 467-472.

9 Forsdyke DR. The liquid scintillation counter as an analogy for the distinction between 'self' and 'not-self' in immunological systems. *Lancet* 1968; **291**: 281-283.

10 Doherty M, Robertson MJ. Some early trends in immunology. *Trends Immunol* 2004; **25**: 623-631.

11 Holland SJ, Bartok I, Attaf M, Genolet R, Luescher IF, Kotsiou E, et al. The T-cell receptor is not hardwired to engage MHC ligands. *Proc Natl Acad Sci USA* 2012; **109**: E3350–e3357.

12 Van Laethem F, Tikhonova AN, Pobezinsky LA, Tai X, Kimura MY, Le Saout C, et al. Lck availability during thymic selection determines the recognition specificity of the T cell repertoire. *Cell* 2013; **154**: 1326-1341.

13 Sprent J, Lo D, Gao E-K, Ron Y. T-cell selection in the thymus. *Immunol Revs* 1988; **101**: 173-190.

14 Detours V, Perelson AS. Explaining alloreactivity as a quantitative consequence of affinity-driven thymocyte selection. *Proc Natl Acad Sci USA* 1999; **96**: 5153-5158.

15 von Boehmer H. Positive and negative selection in Basel. *Nat Immunol* 2008; **9**: 571-573.

16 Forsdyke DR. 'Altered-self' or 'near-self' in the positive selection of lymphocyte repertoires? *Immunol Lett* 2005; **100**: 103-106.

17 Calis JJA, Maybeno M, Greenbaum JA, Weiskopf D, De Silva AD, Sette A, et al. Properties of MHC class I presented peptides that enhance immunogenicity. *PLoS Comput Biol* 2013; **9**: e1003266.

18 Mandl JN, Monteiro JP, Vriskoop N, Germain RN. T cell-positive selection uses self-ligand binding strength to optimize repertoire recognition of foreign antigens. *Immunity* 2013; **38**: 263-274.

19 Cancro MP, Kearney JF. B cell positive selection: road map to the primary repertoire? *J Immunol* 2004; **173**: 15-19.

20 Forsdyke DR. Early evolution of MHC polymorphism. *J Theor Biol* 1991; **150**: 451-456.

21 Forsdyke DR. Adaptive value of polymorphism in intracellular self/not-self discrimination? *J Theor Biol* 2001; **210**: 425-434.

22 Borghans JAM, Beltman JB, de Boer RJ. MHC polymorphism under host-pathogen coevolution. *Immunogenetics* 2004; **55**: 732-739.

23 Calis JJA, de Boer RJ, Kesmir C. Degenerate T-cell recognition of peptides on MHC molecules creates large holes in the T-cell repertoire. *PLOS Comput Biol* 2012; **8**: e1002412.

24 Birnbaum ME, Mendoza JL, Sethi DK, Dong S, Glanville J, Dobbins J, et al. Deconstructing the peptide-MHC specificity of T cell recognition. *Cell* 2014; **157**: 1073-1087.

25 Ernst B, Lee D-S, Chang JM, Sprent J, Surh CD. The peptide ligands mediating positive selection in the thymus control T cell survival and homeostatic proliferation in the periphery. *Immunity* 1999; **11**: 173-181.

26 Persaud SP, Parker CR, Lo WL, Weber KS, Allen PM. Intrinsic CD4+ T cell sensitivity and response to a pathogen are set and sustained by avidity for thymic and peripheral complexes of self peptide and MHC. *Nat Immunol* 2014; **15**: 266-274.

27 Yewdell JW, Reits E, Neefjes J. Making sense of mass destruction: quantitating MHC class 1 antigen presentation. *Nat Rev Immunol* 2003; **3**: 952-961.

28 Apcher S, Daskalogianni C, Lejeune F, Manoury B, Imloos G, Heslop L, et al. Major source of antigenic peptides for the MHC Class 1 pathway is produced during a pioneer round of mRNA translation. *Proc Natl Acad Sci USA* 2011; **108**: 11572-11577.

29 Colbert JD, Farfán-Arribas DJ, Rock KL. Substrate-induced protein stabilization reveals a predominant contribution from mature proteins to peptides presented on MHC class I. *J Immunol* 2013; **191**: 5410-5419.

30 Rock KL, Farfán-Arribas DJ, Colbert JD, Goldberg AL. Re-examining class-I presentation and the DRiP hypothesis. *Trends Immunol* 2014; **35**: 144-152.

31 Bourdetsky D, Schmelzer CEH, Admon A. The nature and extent of contributions by defective ribosome products to the HLA peptidome. *Proc Natl Acad Sci USA* 2014; **111**: E1591-1599.

32 Wolf BJ. Princiotta MF. Processing of recombinant *Listeria monocytogenes* proteins for MHC class I presentation follows a dedicated high-efficiency pathway. *J Immunol* 2013; **190**: 2501-1209.

33 Mitchison NA. Induction of immunological paralysis in two zones of dosage. *Proc R Soc B* 1964; **161**: 275-292.

34 Robbins JH. Human peripheral blood in tissue culture and the action of phytohaemagglutinin. *Experientia* 1963; **20**: 164-168.

35 Forsdyke DR. Role of receptor aggregation in complement-dependent inhibition of lymphocytes by high concentrations of Con-A. *Nature* 1977; **267**: 358-360.

36 Forsdyke DR. Lectin pulses as determinants of lymphocyte activation and inactivation during the first six hours of culture: sequential action of concanavalin-A and complement cause cell lysis. *Can J Biochem* 1980; **58**: 1387-1396.

37 Samanta D, Mukherjee G, Ramagopal UA, Chaparro RJ, Nathenson SG, DiLorenzo TP, et al. Structural and functional characterization of a single-chain peptide-MHC molecule that modulates both naïve and activated CD8+ T cells. *Proc Natl Acad Sci USA* 2011; **108**: 13682-13687.

38 Huang J, Brameshuber M, Zeng X, Xie J, Li Q-j, Chien Y-h, et al**.** A single peptide-MHC complex ligand triggers digital cytokine secretion in CD4+ T cells. *Immunity* 2013; **39**: 846-857.

39 Alexander-Miller MA, Leggatt GR, Sarin A, Berzofsky JA. Role of antigen, CD8, and cytotoxic T lymphocyte (CTL) avidity in high dose antigen induction of apoptosis of effector CTL. *J Exp Med* 1996; **184**: 485-492.

40 Forsdyke DR. Serum factors affecting the incorporation of [3H]thymidine by lymphocytes stimulated by antigen: II. Evidence for a role of complement from studies with heated serum. *Immunology* 1973; **25**: 597-512.

41 Milthorp P, Forsdyke DR. Serum factors affecting the incorporation of [3H]uridine by lymphocytes stimulated by concanavalin-A: studies of the role of complement. *Biochem J* 1973; **132**: 803-812.

42 Forsdyke DR. Serum factors affecting the incorporation of [3H]thymidine by lymphocytes stimulated by antigen: Increased divergence between antigen dose-response curves in heated and control serum with cells from rabbits immunized to increase the proportion of high specificity cells. In: Lucus DO (ed). *Regulatory Mechanisms in Lymphocyte Activation. Proceedings of the 11th Leukocyte Culture Conference.* Academic Press: New York, 1977, pp 730-732.

43 Liblau RS, Tisch R, Shokat K, Yang X-D, Dumont N, Goodnow CC, et al. Intravenous injection of soluble antigen induces thymic and peripheral T-cell apoptosis. *Proc Natl Acad Sci USA* 1996; **93**: 3031-3036.

44 Barton GM, Rudensky AY. Requirement for diverse, low-abundance peptides in positive selection of T cells. *Science* 1999; **283**: 67-70.

45 Shortman K, Egerton M, Spangrude GJ, Scollay R. The generation and fate of thymocytes. *Semin Immunol* 1990; **2**: 3-12.

46 Rothenberg E. Death and transfiguration of cortical thymocytes: a reconsideration. *Immunol Today* 1990; **11**: 116-119.

47 Forsdyke DR. Impaired activation of thymus lymphocytes by PHA. *J Immunol* 1969; **103**: 818-823.

48 Ross JO, Melichara HJ, Au-Yeung BB, Herzmarka P, Weiss, A, Robey EA. Distinct phases in the positive selection of CD8+ T cells distinguished by intrathymic migration and T-cell receptor signaling patterns. *Proc Natl Acad Sci USA* 2014; **111**: E2550-E2558.

49 Struhl K. Transcriptional noise and the fidelity of initiation by RNA polymerase II. *Nat Struct Mol Biol* 2007; **14**: 103-105.

50 Durinovic-Bello I, Gersuk VH, Ni C, Wu R, Thorpe J, Jospe N, et al. Avidity-dependent programming of autoreactive T cells in T1D. *PLOS One* 2014; **9**: e98074.

51 Kyewski B, Klein L. A central role for central tolerance. *Ann Rev Immunol* 2006; **24**: 571-606.

52 Pinto S, Michel C, Schmidt-Glenewinkel H, Harder N, Rohr K, Wild S, et al. Overlapping gene coexpression patterns in human medullary thymic epithelial cells generate self-antigen diversity. *Proc Natl Acad Sci USA* 2013; **110**: E3497-E3505.

53 Gray D, Abramson J, Benoist C, Mathis D. Proliferative arrest and rapid turnover of thymic epithelial cells expressing Aire. *J Exp Med* 2007; **204**: 2521-2528.

54 Mohan JF, Calderon B, Anderson MS, Unanue ER. Pathogenic CD4[+] T cells recognizing an unstable peptide of insulin are directly recruited into islets bypassing local lymph nodes. *J Exp Med* 2013; **210**: 2403-2414.

55 Kyewski B, Haskins K. Editorial overview. *Curr Opin Immunol* 2012; **24**: 67-70.

56 Hedrick SM. Dawn of the hunt for non-classical MHC function. *Cell* 1992; **70:** 177-180.

57 Forsdyke DR, Madill CA, Smith SD. Immunity as a function of the unicellular state: implications of emerging genomic data. *Trends Immunol* 2002; **23**: 575-579.

58 Forsdyke DR. Heat shock proteins defend against intracellular pathogens: a non-immunological basis for self/not-self discrimination. *J Theor Biol* 1985; **115**: 471-473.

59 Forsdyke DR. Two signal model of self/not-self immune discrimination: an update. *J Theor Biol* 1992; **154**: 109-118.

60 Forsdyke DR. Relationship of X chromosome dosage compensation to intracellular self/not-self discrimination: a resolution of Muller's paradox? *J Theor Biol* 1994; **167**: 7-12.

61 Cristillo AD, Mortimer JR, Barrette IH, Lillicrap TP, Forsdyke DR. Double-stranded RNA as a not-self alarm signal: to evade, most viruses purine-load their RNAs, but some (HTLV-1, Epstein-Barr) pyrimidine-load. *J Theor Biol* 2001; **208**: 475-491.

62 Murat P, Zhong J, Lekieffre L, Cowieson NP, Clancy JL, Preiss T, et al. G-quadruplexes regulate EBV-encoded nuclear antigen 1 mRNA translation. *Nat Chem Bio* 2014; **10**: 358-364.

63 Lauffer MA. *Entropy-driven processes in b*iology. New York: Springer-Verlag; 1975.

64 Rajan RS, Illing ME, Bence NF, Kopito RR. Specificity in intracellular protein aggregation and inclusion body formation. *Proc Natl Acad Sci USA* 2001; **98**: 13060-13065.

65 Smith-Bouvier DL, Divekar AA, Sasidhar M, Du S, Tiwari-Woodruff SK, King JK, et al. A role for sex chromosome complement in the female bias in autoimmune disease. *J Exp Med* 2008; **205**: 1099-1108.

66 Forsdyke DR. X-chromosome reactivation perturbs intracellular self/not-self discrimination. *Immunol Cell Biol* 2009; **87**: 525-528.

67 Dillon SP, Kurien BT, Li S, Bruner GR, Kaufman KM, Harley JB, et al. Sex chromosome aneuploidies among men with systemic lupus erythematosus. *J Autoimmun* 2012; **38**: J129-J134.

68 Forsdyke DR. Heat shock proteins as mediators of aggregation-induced 'danger' signals: Implications of the slow evolutionary fine-tuning of sequences for the antigenicity of cancer cells. *Cell Stress Chaperones* 1999; **4**: 205-210.

69 Mosenson JA, Zloza A, Klarquist J, Barfuss AJ, Guevara-Patino JA, Poole IC. HSP70i is a critical component of the immune response leading to vitiligo. *Pigment Cell Melanoma Res* 2012; **25**: 88-98.

70 Bersuker K, Hipp MS, Calamini B, Morimoto RI, Kopito RR. Cellular model of Huntington disease exacerbates inclusion body formation in a heat shock response activation. *J Biol Chem* 2013; **288**: 23633-23638.

71 Forsdyke DR. Functional constraint and molecular evolution. In: *Encyclopedia of Life Sciences (eLS)*. Wiley; Chichester, 2012.

72 McConkey AH. Molecular evolution, intracellular organization, and the quinary structure of proteins. *Proc Nat Acad Sci USA* 1982; **79**: 3236-3240.

73 Vogel C. Protein expression under pressure. Science 2013; **342**: 1052-1053.

74 Hoof I, Baarle D van, Hildebrand WH, Kesmir C. Proteome sampling by the HLA Class I antigen processing pathway**.** *PLOS Comput Biol* 2012; **8**: e1002517.

75 Gidalevitz T, Ben-Zvi A, Ho KH, Brignull HR, Morimoto RI. Progressive disruption of cellular protein folding in models of polyglutamine diseases. *Science* 2006; **311**: 1471-1474.

76 Fortier M-H , Caron E, Hardy M-P, Voisin G, Lemieux S, Perreault C , et al. The MHC class I peptide repertoire is molded by the transcriptome. *J Exp. Med* 2008; **205**: 595-610.

77 Mommen GPM, Frese CK, Meiring HD, Brink J van G van den, Jong APJM de, Els CACM van, et al. Expanding the detectable HLA peptide repertoire using electron-transfer/higher-energy collision dissociation. *Proc Natl Acad Sci USA* 2014; **111**: 4507-4712.

78 Yang J-R*,* Liao B-Y*,* Zhuang S-M*,* Zhang J*.* Protein misinteraction avoidance causes highly expressed proteins to evolve slowly. *Proc  Nat  Acad  Sci USA* 2012; **109**: E831-E840*.*

79 Levy ED, De S, Teichmann SA. Cellular crowding imposes global constraints on the chemistry and evolution of proteins. *Proc Nat Acad Sci USA* 2012; **109**: 20461-20466.

80 Tartaglia GG, Pechmann S, Dobson CM, Vendruscolo M. Life on the edge: a link between gene expression levels and aggregation rates of human proteins. *Trends Bioch Sci* 2007; **32**: 204-206.

81 Li H-Y, Yeh P-A, Chiu H-C, Tang C-Y, Tu BP-h. Hyperphosphorylation as a defense mechanism to reduce TDP-43 aggregation. *PLoS One* 2011; **6**: e23075.

82 Gong B, Kielar C, Morton AJ. Temporal separation of aggregation and ubiquitination during early inclusion formation in transgenic mice carrying the Huntington's disease mutation. *PLoS One* 2012; **7**: e41450.

83 Forsdyke DR. *Evolutionary bioinformatics*. 2[nd] ed. Springer; New York, 2011, pp 295-337.

84 Villasenor J, Besse W, Benoist C, Mathis D. Ectopic expression of peripheral-tissue antigens in the thymic epithelium: probabilistic, monoallelic, misinitiated. *Proc Natl Acad Sci USA* 2008; **105**: 15854-15859.

85 Ashton-Rickardt PG, Tonegawa S. A differential avidity model for T-cell selection. *Immunol Today* 1994; **15**: 362-366.

86 Sebzda E, Wallace VA, Mayer J, Yeung RS, Mak TW, Ohashi PS. Positive and negative thymocyte selection induced by different concentrations of a single peptide. *Science* 1994; **263**: 1615-1618.

87 Bains I, van Santen HM, Seddon B, Yates AJ. Models of self-peptide sampling by developing T cells identify candidate mechanisms of thymic selection. *PLoS Comput Biol* 2013; **9**: e1003102.

88 Suzuki H, Guinter TI, Koyasu S, Singer A. Positive selection of CD4+ T cells by TCR-specific antibodies requires low valency TCR cross-linking: implications for repertoire selection in the thymus. *Eur J Immun* 1998; **28**: 3252-3258.

89 Girao C, Hu Q, Sun J, Ashton-Rickardt PG. Limits on the differential avidity model of T cell selection in the thymus. *J Immunol* 1997; **159**: 4205-4211.

90 Klein L, Hinterberger M, Wirnsberger G, Kyewski B. Antigen presentation in the thymus for positive selection and central tolerance induction. *Nat Rev Immunol* 2009; **9**: 833-844.

91 Manz BN, Jackson BL, Petit RS, Dustin ML, Groves J. T-cell triggering thresholds are modulated by the number of antigen within individual T-cell receptor clusters. *Proc Natl Acad Sci USA* 2011; **108**: 9089-9094.

92 Bosch B, Heipertz EL, Drake JR, Roche PA. Major histocompatibility complex (MHC) class II-peptide complexes arrive at the plasma membrane in cholesterol-rich microclusters. *J Biol Chem* 2013; **288**: 13236-13242.

93 Cho J-H, Kim H-O, Surh CD, Sprent J. T cell receptor-dependent regulation of lipid rafts controls naïve CD8+ T cell homeostasis. *Immunity* 2010; **32**: 214-226.

94 Forsdyke DR, Mortimer JR. Chargaff's legacy. *Gene* 2000; **261**: 127-137.

95 Forsdyke DR. Implications of HIV RNA structure for recombination, speciation, and the neutralism-selectionism controversy. *Microb Infect* 2014; **16:** 96-103.